\documentclass[final,1p,times,twocolumn]{elsarticle}

\usepackage{amssymb}
\usepackage{amsthm}
\usepackage[nolist,printonlyused]{acronym}
\usepackage{hyperref,graphicx}
\usepackage[utf8]{inputenc} 

\journal{New Astronomy}

\hypersetup{colorlinks=true, citecolor=blue, linkcolor= magenta}
\usepackage{booktabs}

\newcommand{\apj}{ApJ}
\newcommand{\apjl}{ApJ}
\newcommand{\apjs}{ApJS}

\newcommand{\mnras}{MNRAS}
\newcommand{\nat}{Nature}
\newcommand{\nar}{New Astronomy Reviews}

\newcommand{\aapr}{A{\&}ARv}

\newcommand{\pasj}{PASJ}

\begin{acronym}[TDMA]
 \acro{LMXB}{low mass X-ray binary}
 \acrodefplural{LMXB}{Low mass X-ray binaries}
 \acro{AMXP}{accreting millisecond X-ray pulsar}
 \acro{RXTE/PCA}{proportional counter array}
 \acro{RXTE}{the Rossi X-ray timing explorer}
 \acro{MAXI}{the monitor of all sky X-ray image}
 \acro{HMXB}{high mass X-ray binary}
 \acrodefplural{HMXB}{High mass X-ray binaries}
 \acro{SWIFT/XRT}{the Swift gamma-ray burst mission/X-ray telescope}
 \acro{XSPEC}{an X-ray spectral fitting package}
 \acro{FRED}{fast-rise-exponential-decay}
 \acro{DIM}{the disc instability model}
 \acro{LIS}{low-intensity-state}
 \acro{SXT}{soft X-ray transient}
 \acro{HS}{high/soft state}
 \acro{LH}{low/hard state}
 \acro{ASM}{the all-sky monitor}
 \acro{PCU}{proportional counter unit}
 \acro{ISS}{international space station}
 \acro{MSP}{millisecond pulsar}
 \acro{NS}{neutron star}
\end{acronym}

\begin{document}

\begin{frontmatter}



\title{On the Morphology of Outbursts of Accreting Millisecond X-ray Pulsar Aquila~X-1}


\author{C.\ G\"{u}ng\"{o}r$^{1}$, K.\ Y.\ Ek\c{s}i$^{2}$, \& E.\ G\"o\u{g}\"u\c{s}$^{1}$}
\address{$^{1}$Sabanc{\i} University, Faculty of Engineering and Natural Science, Orhanl{\i} $-$ Tuzla, 34956, \.{I}stanbul, Turkey\\
$^{2}$\.{I}stanbul Technical University, Faculty  of Science  and  Letters,  Physics Engineering  Department,
  34469,  \.{I}stanbul, Turkey \\}
\ead{gungorcan@itu.edu.tr}
\begin{abstract}

We present the X-ray light curves of the last two outbursts --2014 \& 2016-- of the well known 
\ac{AMXP} Aquila X-1 using \ac{MAXI} observations in the \hbox{$2-20$~keV} band.
After calibrating the \ac{MAXI} count rates to \ac{ASM} level, we 
report that the 2016 outburst is the most energetic event of Aql~X-1, ever observed from this source. 
We show that 2016 outburst is a 
member of the \textit{long-high} class according to the classification presented by 
G\"{u}ng\"{o}r et al. with $\sim68$~cnt/s maximum flux 
and $\sim60$~days duration time and the previous outburst, 2014, belongs to the \hbox{\textit{short-low}} class with
$\sim 25$~cnt/s maximum flux and $\sim 30$~days duration time.
In order to understand differences between outbursts,
we investigate the possible dependence of the peak intensity to the quiescent duration leading to the outburst and
find that the outbursts following longer quiescent episodes tend to reach higher peak energetic.

\end{abstract}
\end{frontmatter}


\section{Introduction}
\label{intro}

Aql X-1, discovered by \citet{kunte73} in 1973, is a \ac{LMXB} in which a neutron star accretes matter from a disk fed by its K-type companion via Roche lobe overflow \citep{fra+02}. It displays thermonuclear X-ray bursts \citep{koy+81} every few hours due to accumulation of matter on its surface. The burst oscillations \citep{zha+98bo} as well as the measured spin frequency of $550.273$~Hertz (spin period is $1.8$~msec)  \citep{cas+08} indicate to a rapidly spinning neutron star likely spun up in accordance with the recycling hypothesis \citep{alp+82}. The detection of pulsations only during a limited episode indicates the object is an intermittent  \ac{AMXP}.

Aql X-1 is also classified as a \ac{SXT} \citep[see][for a review]{cam+98rev} as it shows outbursts almost each year in its X-ray light curve due to the thermal-viscous instability in the accretion disk \citep[see][for a review]{las01}.
We have a wealth of data of these outbursts thanks to the \ac{ASM} aboard \ac{RXTE} which monitors the source since 1996 until the end of the mission, and to the \ac{MAXI} aboard \ac{ISS} for ongoing observations since 2009. These continuous observations allow Aql X-1 to be a suitable source for studying the outbursts of \acp{SXT}.

The morphology of the outbursts of Aql X-1 has been studied by \citet{mai08} via optical and near-infrared 
observations. They identify two types of events; the \ac{FRED} type outbursts that are mostly explained with \ac{DIM} \citep{las01, che+97}
and the \ac{LIS} events in which the structures of these  outburst are in a more complicated variable flux state. 
which does not exceed the 5 cnt/s level and can last longer than a month.

A broad classification of the \ac{FRED} type outbursts of Aql X-1 is presented by \citet{gungor+14} (hereafter G14)
who showed that \ac{FRED} type outbursts of Aql~X-1 exhibit three main classes depending on the peak flux and the outburst duration: the \textit{short-low}, the \textit{medium-low} and the \textit{long-high} outbursts. 
The underlying physical cause of the differences between these classes is still unclear.

We present the X-ray light curves of the 2014 and the 2016 outbursts 
in the light of the classification of G14 in \autoref{method}. 
We explain, in that section, the procedures that we followed to define the 
outbursts and the durations of the quiescent stages, and the relation between them.
We discuss and present the conclusions of our work in \autoref{discuss}.

\begin{figure*}[ht]
\centering
  \includegraphics[angle=0,width=0.9\textwidth]{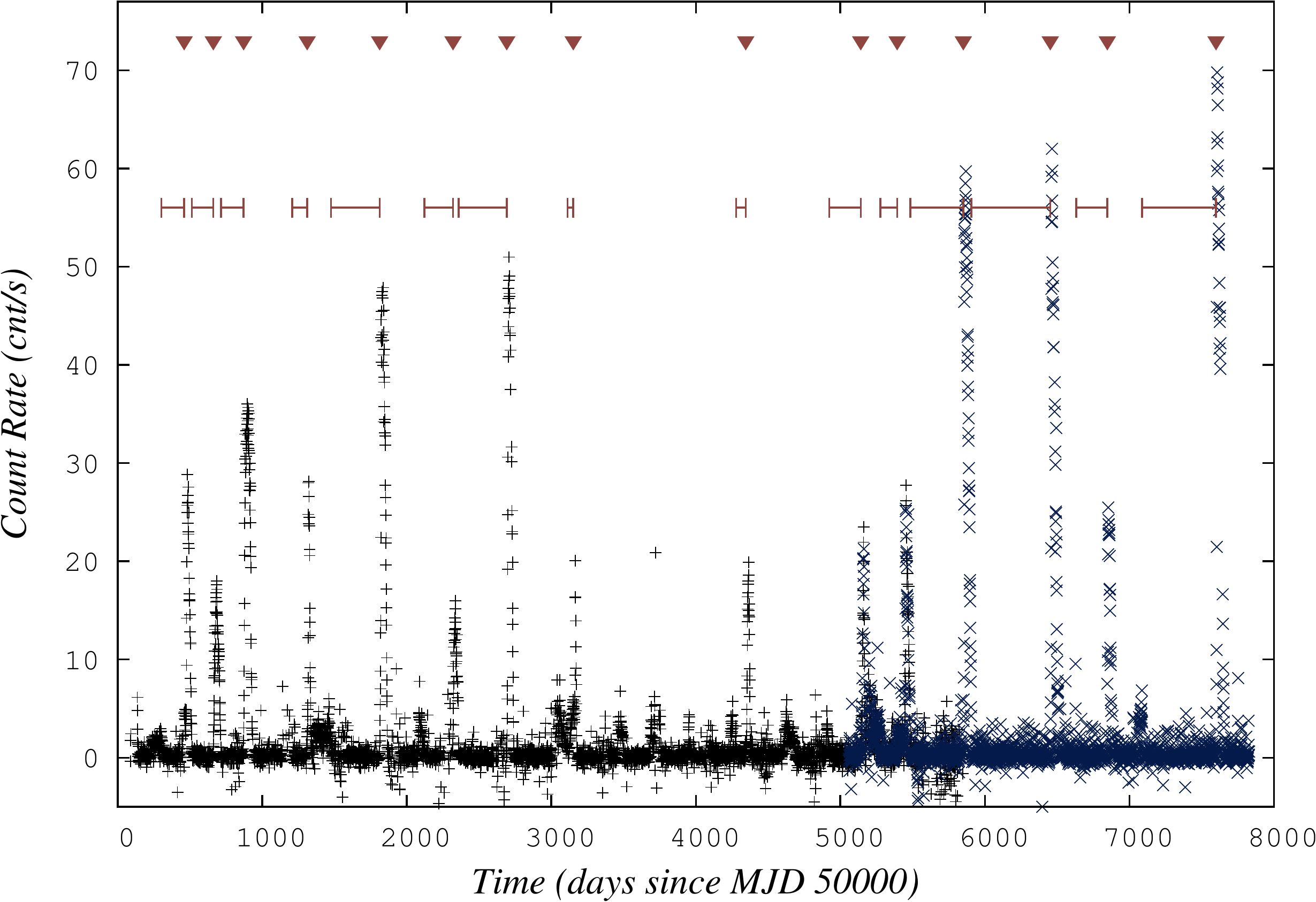}
     \caption{21 years light curve of Aql X-1 since 1996.
     The black pluses and the dark blue crosses represent the data obtained from the \ac{ASM} and the \ac{MAXI}, respectively. 
     The brown upside-down triangles show the \ac{FRED} type outbursts that are used in the classification study.
     The pre-outburst quiescent period for each outburst is indicated with the brown lines.}
         \label{all}
\end{figure*}

\section{Methodology and Results}
\label{method}

We, first, obtained all daily average fluxes from the \ac{MAXI} \citep{mat+09} and the \ac{ASM} in the energy range of 
\hbox{$1.3-12.1$~keV} and \hbox{$2-20$~keV}, respectively. These bands are the largest ranges for each detector.
Following G14, we calibrated the \ac{MAXI} data with the \ac{ASM} data using the peak count rate of the 2009 and the 2010 
outbursts which were observed by both detectors.
In \autoref{all}, we present the long term light curve of Aql X-1 displaying all the outburst of the source since 1996.

\begin{figure}[t]
\centering
  \includegraphics[angle=0,width=0.9\textwidth]{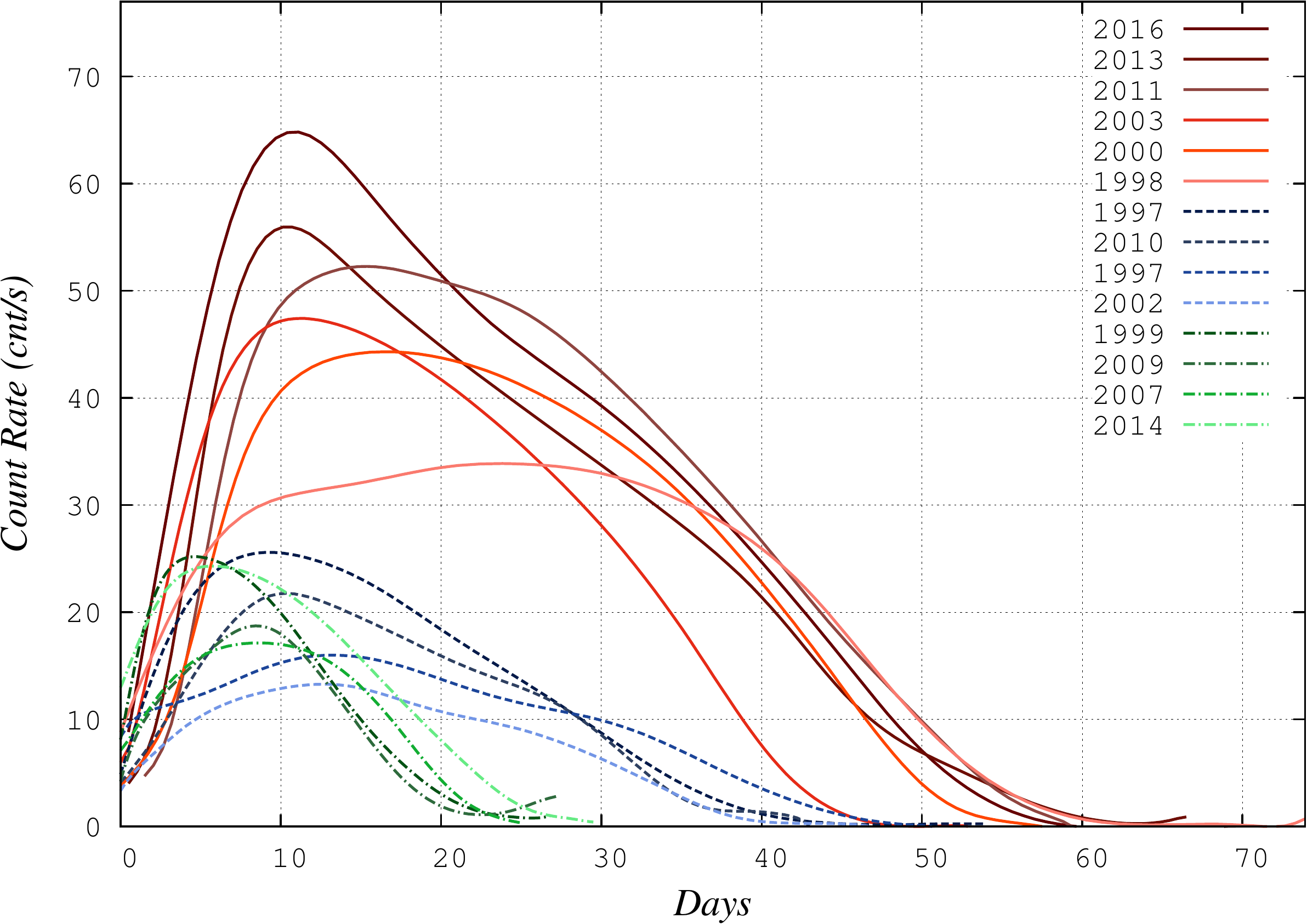}
     \caption{The smoothed light curves of the outbursts of Aql~X-1, calibrated based on
     the beginnings of the outbursts. This figure is an updated version of the 
     one presented by G14 with the addition of the 2014 and 2016 outbursts.
     The labels from the top downwards represent the peak fluxes in 
     order of maximum brightness.}
         \label{light}
\end{figure}

We, then smoothed the light curves of the 2014 and the 2016 outbursts with a
``natural'' spline formalism  \citep{spline} as done for the earlier outbursts in G14.
In \autoref{light}, we present the set of Aql X-1 outburst morphologies.
Correspondingly, based on the outburst classification scheme of G14, we find that the 2014 and the 2016 outbursts fit into the \textit{short-low} 
and \textit{long-high} types, respectively.

We also investigated the possible relation between outburst characteristics and time passed
prior to the onset of the activity. The essential step in this investigation was to
establish a scheme to define the onset and the end of outbursts. We first selected
time intervals with no activity to determine the average count rate and its standard
deviation for the quiescent level. Assuming the $3\sigma$ level above quiescence as 
the threshold to identify a physical change in the light curve, we mark the beginning 
and the end of outbursts as the first and the last excess above the threshold level,
respectively. We also require an outburst episode to have at least five individual
measurement for a reliable identification. The brown horizontal lines in \autoref{all} 
indicate the intervals of pre-outburst quiescent stages for \ac{FRED} type outbursts,
and the upside-down triangles mark the onset times of these outbursts.

\begin{figure*}[ht]
\centering
  \includegraphics[angle=0,width=0.9\textwidth]{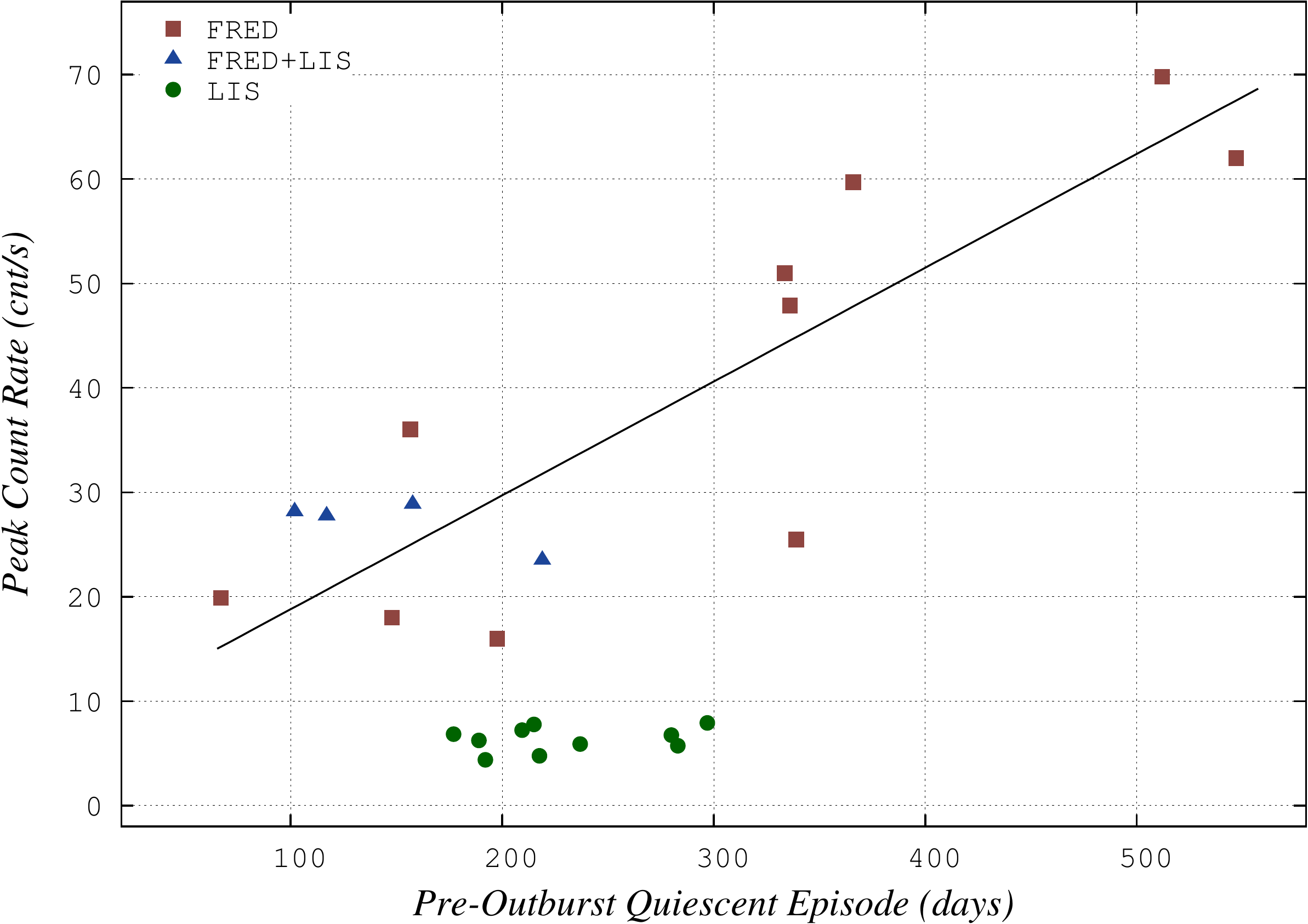}
     \caption{The relation between the peak fluxes of the outbursts and the  durations of the pre-outburst quiescent episodes.
     The brown squares, the blue triangles and the green circles represent the \ac{FRED} type outburst,
     the \ac{FRED} type outbursts with \ac{LIS}
     companion and the \ac{LIS} type outbursts, respectively.
     }
         \label{time}
\end{figure*}

We find that the maximum intensity of Aql X-1 outbursts is positively correlated with
the length of quiescent episode prior to that particular outburst. We present this
correlation with square symbols in \autoref{time}. Quantitatively, we obtain a
Spearman's rank order correlation coefficient of 0.81 with the chance probability of 
$4.8x10^{-3}$ for the correlation between pre-outburst quiescent duration and the 
peak flux for \ac{FRED} type of outbursts. We also fit the peak intensity vs. waiting
time trend of these types outbursts with a first order polynomial, which yields a minimum
peak rate of 7.9 counts/s and the slope of 0.109$\pm$0.023. Note that such a correlation
is not the case for the much lower intensity LIS type nor for FRED+LIS type of 
outbursts.

\section{Discussion \& Conclusion}
\label{discuss}

In this study, we updated the classification introduced in G14 by adding 
the latest two outbursts --2014 \& 2016-- (\autoref{light}) of Aql~X-1 to the list.
We showed that the outburst starting at July 2014 with almost $30$~days duration 
and $25$~cnt/s maximum flux is a member of the
short-low type, and the outburst starting at July 2016 with $60$~days
duration and $68$~cnt/s maximum flux is a member of 
the long-high type. This lends credit to the view that the classification scheme introduced by  G14 is robust.

The long-term evolution of the X-ray flux of Aql X-1 shows that the energy released varies from one outburst to another. 
Although it displays at least one outburst almost each year, we see that the system passed through a relatively quiet episode 
between 2003 June and 2011 December during when it showed no outburst brighter than $30$~cnt/s (\autoref{all}), but
exhibited many \ac{LIS} type low energetic events.
This propounds that the 
accretion reservoir is diminishing relatively calmly resulting in low-energetic events rather than leading to 
\ac{FRED} type outbursts.

To explore the underlying cause for the differences between outburst types, 
we searched for a relation between the maximum intensity of the outbursts and the durations of the preceding quiescent episodes.
We considered both the \ac{FRED} and the \ac{LIS} type events to identify the active and quiescent stages, therefore,
we took into account the released energy even via weaker events. 
Thereby, unlike \citet{cam+13}, we introduce possible relation between peak fluxes of the outbursts and the durations of the 
preceding quiescent episodes as can be seen from \autoref{time}.
The scattered nature of the data in \autoref{time} around the linear fit implies 
that the waiting time is not the only parameter determining the energy to be released in the forthcoming outburst.
On the other hand, we observe that the peak fluxes do slightly increase with the durations of the quiescent episodes. 
In this case, the longer waiting time might lead to accumulation of more material in the disc 
resulting in a more luminous outburst.

Finally, using the ratio of the peak count rates of the 2016 and 1997 outbursts,
and the given peak luminosity of the 1997 outburst in \citet{cam+14},
the peak luminosity of the brightest outburst of Aql X-1 in 2016 is estimated 
as $\sim8.7\times10^{37}$erg/s (the distance of the source is 4.5 kpc \citep{gal+08}). 
This corresponds to a peak luminosity $\sim0.5$ L$_{Edd}$ for a $1.4$~M$_\odot$ neutron star accretor,
after about 512 days of quiescence.

\section{Acknowledgement}

We thank the anonymous referee for constructive comments.
This research has made use of the MAXI data provided by RIKEN, JAXA and the MAXI team 
and the results provided by the ASM/RXTE teams at MIT and at the RXTE SOF and GOF at NASA's GSFC.


\end{document}